\documentclass[aps,floatfix,twocolumn,prl]{revtex4}

\usepackage{epsfig,amsmath,graphicx,graphics,float}

\begin{document}

\title{Discontinuities in the First and Second Sound Velocities at the Berezinskii-Kosterlitz-Thouless Transition}

\author{Tomoki Ozawa and Sandro Stringari}
 
\affiliation{
INO-CNR BEC Center and Dipartimento di Fisica, Universit\`a di Trento, I-38123 Povo, Italy
}%

\date{\today}

\begin{abstract}
We calculate the temperature dependence of the first and second sound velocities in the
superfluid phase of a 2D dilute Bose gas by solving Landau's two fluid hydrodynamic equations. We predict the occurrence of a
significant discontinuity in both velocities at the critical temperature, as a
consequence of the jump of the superfluid density characterizing
the Berezinskii-Kosterlitz-Thouless transition. The key role of the
thermal expansion coefficient is discussed. We find that second sound in
this dilute Bose gas can be easily excited through a density perturbation,
thereby, making the perspective of the measurement of the superfluid
density particularly favorable.
\end{abstract}

\maketitle

Two dimensional superfluids differ in a profound way from their 3D counterparts. In fact, the Hohenberg-Mermin-Wagner theorem~\cite{Hohenberg1967, Mermin1966} rules out the occurrence of long range order at finite temperature in 2D systems with continuous symmetry. Furthermore, the superfluid density approaches a finite value at the critical point of a 2D superfluid, known as Berezinskii-Kosterlitz-Thouless (BKT) transition~\cite{Berezinskii1972, Kosterlitz1972, Nelson1977}, rather than vanishing, as happens in 3D.
With respect to the second order phase transitions characterizing the
onset of superfluidity in 3D, the nature of the BKT phase transition is
deeply different, being associated with the
emergence of a topological order, resulting from the pairing of
vortices and antivortices.
A peculiar property of these 2D systems is also the absence of discontinuities in the thermodynamic functions at the critical temperature characterizing the transition to the superfluid phase. In order to identify the transition point, one has consequently to measure suitable transport properties. This is the case of the recent experiment of Ref.~\cite{Desbuquois2012} on dilute two dimensional Bose gases where the superfluid critical velocity was measured in a useful range of temperatures, pointing up the occurrence of a sudden jump at a critical temperature when one enters the superfluid regime.

In this Letter, we discuss the behavior of both first and second sound in
2D superfluid gases, with particular emphasis on the discontinuity of
their velocities at the critical point, caused by the jump of the
superfluid density.
While first sound corresponds to a wave where the
superfluid and normal components of the fluid move in phase, in the
propagation of second sound, they move with opposite phase. The
experimental measurement of second sound has recently proven successful in determining the temperature dependence of $n_s$ in a 3D strongly interacting Fermi gas~\cite{Sidorenkov2013}
, and our analysis provides an analogous way to determine the superfluid density in 2D Bose gases.
Although the thermodynamics in 2D dilute Bose gases is now well understood both theoretically~\cite{Prokof'ev2001, Prokof'ev2002, Rancon2012} and experimentally~\cite{Hung2011, Yefsah2011}, the measurement of the superfluid density remains one of the main issues of interest. This issue is particularly crucial as the observation of the superfluid jump at the transition would provide a complete proof of the BKT transition.
Besides the prospect of measuring the superfluid density, the measurement of second sound itself would open a new page in the hydrodynamics of 2D superfluids, because
second sound has never been measured in any 2D system so far. In Helium films, the normal component of the liquid is in fact clamped to the  substrate  and cannot participate in the propagation of sound, only the superfluid being  free to move (third sound).
In Helium films, the value of $n_s$ became accessible via  third sound~\cite{Rudnick1978}  and  torsional oscillator measurements~\cite{Bishop1978}, confirming the superfluid jump at the transition.
Since  the trapping of dilute atomic gases is provided by smooth potentials, second sound is expected to propagate in these systems also in 2D and to be properly described by the two fluids Landau's hydrodynamic equations. Its propagation in dilute Bose gases is  expected to exhibit very peculiar features as compared to less compressible fluids like helium or strongly interacting Fermi superfluid gases. In fact, in these latter systems  second sound can be identified as an entropy oscillation and corresponds with good accuracy to an isobaric oscillation~\cite{Hou2013}.  This is not the case of dilute Bose gases which are highly compressible,  giving rise to sizable coupling effects between density and entropy oscillations.  Furthermore, with respect to the 3D case, in 2D, the superfluid density exhibits a jump at the transition and this shows up as a discontinuity of both first and second sound velocities as we will discuss in this Letter.

We start our investigation by considering the Landau's two fluid hydrodynamic equations to describe the dynamics of the system in the superfluid phase of a 2D uniform configuration. The third direction is assumed to be blocked by a tight harmonic confinement, a condition  well achieved in current experiments.
In the absence of  additional confinement in the 2D plane,
Landau's equations provide the following equation for the velocity of sound $c$ at temperature $T$~\cite{PitaevskiiStringari}:
\begin{align}
	c^4
	- \frac{T}{m}\left[ \frac{1}{nT \kappa_s} + \frac{\bar{s}^2 n_s}{\bar{c}_v n_n}\right]c^2
	+ \frac{T^2}{m^2}\frac{\bar{s}^2 n_s}{\bar{c}_v n_n}\frac{1}{nT \kappa_T}
	=
	0, \label{landaueq}
\end{align}
where $m$ is the mass of the constituents and all the thermodynamic quantities should be calculated in 2D:
the specific heat at constant volume $\bar{c}_v$, the entropy density $\bar{s}$, the superfluid and normal densities $n_s$ and $n_n$, the isothermal and isoentropic compressibilities $\kappa_T \equiv (\partial n/ \partial p)_T / n$ and $\kappa_s \equiv (\partial n/ \partial p)_s / n$, where $s$ is the entropy.
We use the units $\hbar = k_B = 1$ throughout this Letter.
There are two nonnegative solutions of Eq.~(\ref{landaueq}); the first and second sound velocities are, respectively,  the larger and the smaller ones.

In the following, we focus on the dilute 2D Bose gas, where all the thermodynamic ingredients can be written in terms of universal dimensionless functions~\cite{Prokof'ev2002, Hung2011, Yefsah2011}. These depend only on the variable $x \equiv \mu/T$ and on the dimensionless coupling constant $g = \sqrt{8\pi}a/l$, where $\mu$ is the chemical potential, $a$ is the three-dimensional scattering length, and $l$ is the oscillator length in the confined direction.
Here, we assume $l \gg a$ so that the interaction is momentum independent~\cite{Petrov2000}.
We introduce the dimensionless reduced pressure $\mathcal{P}$ and the phase space density $\mathcal{D}$ by
\begin{align}
	\mathcal{P}(x, g) &\equiv \lambda_T^2 P/T,
	&
	\mathcal{D}(x,g) &\equiv \lambda_T^2 n,
\end{align}
where $\lambda_T \equiv \sqrt{2\pi/mT}$ is the thermal de Broglie wavelength, $P$ is the ordinary pressure, and $n$ is the particle number density. The  simple relation $\partial \mathcal{P}/\partial x = \mathcal{D}$ follows from thermodynamics. Straightforward thermodynamic calculations lead to
\begin{align}
	\bar{s}
	&=
	2\frac{\mathcal{P}}{\mathcal{D}} - x,
	&
	\kappa_T
	&=
	\frac{1}{nT}\frac{\mathcal{D}^\prime}{\mathcal{D}},
	\hspace{0.6cm}
	\kappa_s
	=
	\frac{1}{nT}\frac{\mathcal{D}}{2\mathcal{P}},
	\notag \\
	\bar{c}_v
	&=
	2\frac{\mathcal{P}}{\mathcal{D}} - \frac{\mathcal{D}}{\mathcal{D}^\prime},
	&
	\bar{c}_p
	&=
	\left( 2\frac{\mathcal{P}}{\mathcal{D}} - \frac{\mathcal{D}}{\mathcal{D}^\prime} \right)
	2\frac{\mathcal{P}\mathcal{D}^\prime}{\mathcal{D}^2},
\end{align}
where $\mathcal{D}^\prime \equiv \partial \mathcal{D}/\partial x$, and $\bar{c}_p$ is the specific heat at constant pressure.
The universal function $\mathcal{D}$ has been numerically calculated for small $g$ in~\cite{Prokof'ev2002}, and both $\mathcal{P}$ and $\mathcal{D}$ have been theoretically~\cite{Rancon2012} and experimentally~\cite{Hung2011, Yefsah2011} determined around the superfluid transition. The results available  from different methods well agree with each other.

The superfluid  density $n_s$ cannot be calculated in terms of the  universal functions introduced above, but can be nevertheless expressed in terms of another dimensionless function $\mathcal{D}_s (x,g) \equiv \lambda_T^2 n_s$, which is known close to the transition in~\cite{Prokof'ev2002} as well as in the highly degenerate phonon regime (large and positive $x$). (The temperature dependence of the superfluid density was also calculated when the interaction is strong, beyond the present weakly interacting regime, using quantum Monte Carlo methods~\cite{Pilati2008}.) At the critical point, $\mathcal{D}_s=4$ following from the universal result $n_s = 2mT_c/\pi$ \cite{Nelson1977}, where $T_c$ is the BKT transition temperature, providing the relationship between the jump of the superfluid density and the critical temperature in 2D superfluids.

The superfluid transition of a weakly interacting gas is predicted to take place at the value ~\cite{Prokof'ev2001}
\begin{align}
	x_c = \frac{g}{\pi} \log \left( \frac{\xi_\mu}{g} \right),
\end{align}
where $\xi_\mu \approx 13.2$.
For example, for $g = 0.1$, a value relevant for the experiments of~\cite{Yefsah2011, Tung2010, Desbuquois2012}, the critical point corresponds to $x_c \approx 0.16$ or, in terms of the density, to  $ T_c = \{2\pi/\mathcal{D}(x_c)\}n/m \approx 0.76 n/m$.

\begin{figure}[htbp]
\begin{center}
\includegraphics[width=7.0cm]{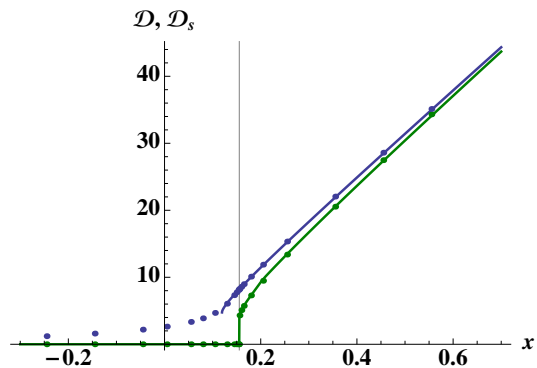}
\caption{Phase space density $\mathcal{D}(x)$ (upper line) and superfluid density $\mathcal{D}_s (x)$ (lower line), calculated from the approximate analytical formula given in~\cite{Prokof'ev2002} for $g=0.1$. The numerical values given in the same reference are also plotted as dots. The transition $x_c \approx 0.16$ is marked as a vertical line. For large $x$, corresponding to $T/T_c \to 0$ limit, both functions approach $\sim 2\pi x/g$.}
\label{dvx}
\end{center}
\end{figure}

In~\cite{Prokof'ev2002}, approximate analytical formula for $\mathcal{D}$ and $\mathcal{D}_s$, as well as their numerical values, are given around the transition, and it was found that the analytical formula agree well with the numerical values. In the following  we use the analytical formula for $\mathcal{D}$ and $\mathcal{D}_s$ to calculate the relevant thermodynamic quantities. We determine $\mathcal{P}$ by integrating $\mathcal{D}$ with respect to $x$, with the constant of the integral chosen so that the reduced pressure at the transition coincides with the value calculated from the Hartree-Fock mean-field theory described in~\cite{Yefsah2011}. The resulting value of $\mathcal{P}$  is consistent  with the experimental result of~\cite{Yefsah2011}.
The dimensionless functions $\mathcal{D}$ and $\mathcal{D}_s$ are plotted in Fig~\ref{dvx} for $g = 0.1$.
The approximate analytical formula are not valid at high temperatures (negative  $x$), corresponding to $T \gg T_c$.
In the figure, the numerical values given in~\cite{Prokof'ev2002} are also plotted. From these numerical values, $\mathcal{D}^\prime$ can also be obtained by using finite difference method. The numerical values of $\mathcal{P}$ is not presently available, and thus, we take the values from approximate analytical expression mentioned above when necessary.

\begin{figure}[htbp]
\begin{center}
\includegraphics[width=7.0cm]{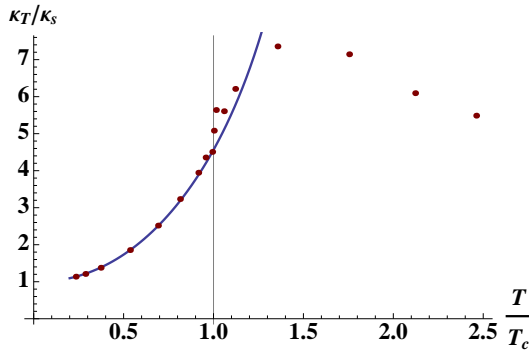}
\caption{Value of $\kappa_T / \kappa_s = \bar{c}_p / \bar{c}_v = 2\mathcal{P}\mathcal{D}^\prime/\mathcal{D}^2$ as a function of $T/T_c$ for $g = 0.1$. The line is calculated using the approximate analytical expressions on the universal functions, and dots are calculated using the numerical results from~\cite{Prokof'ev2002}.}
\label{ktksratio}
\end{center}
\end{figure}

\begin{figure}[htbp]
\begin{center}
\includegraphics[width=7.0cm]{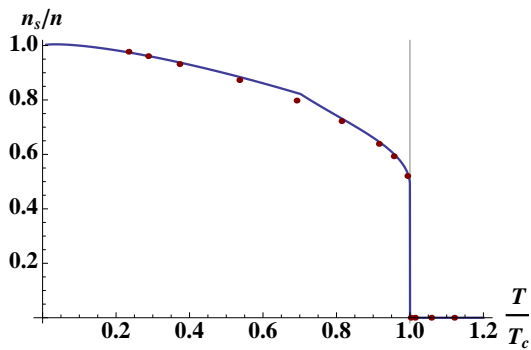}
\caption{Normalized superfluid density $n_s/n$ for $g = 0.1$. The line is calculated from the approximate analytical expression, and the dots are calculated using the numerical values from~\cite{Prokof'ev2002}. Two analytical expressions valid at low and high temperatures are connected to give the curve, resulting in an unphysical kink at $T/T_c \sim 0.7$.}
\label{nslines}
\end{center}
\end{figure}

In Figs.~\ref{ktksratio} and~\ref{nslines}, we show two relevant thermodynamic functions: the ratio $\kappa_T/\kappa_s = \bar{c}_p/\bar{c}_v$ and the superfluid fraction $n_s/n$, calculated as a function of $T/T_c$ for a fixed value of the total density. The figures correspond to the value $g = 0.1$. 
Figure~\ref{ktksratio} clearly points out the large difference between $\kappa_T$ and $\kappa_S$ near the critical point, reflecting the large value of the thermal expansion coefficient $\alpha \equiv - (\partial n/\partial T)_p/n = T^{-1}(\kappa_T/\kappa_s - 1)$.
We note that $\kappa_T/\kappa_s \to 1$ as $T \to 0$ while  $\kappa_T/\kappa_s \to 2$ as $T \to \infty$.
The figure shows the occurrence of a maximum above the critical point, whose height becomes larger and larger as one decreases the value of $g$.
Figure~\ref{nslines} instead points out the large value of the superfluid fraction at the transition and the consequent jump. As we will see below, both the large value  of the thermal expansion coefficient and the jump of the superfluid density play an important role to characterize the solutions of the Landau's Eq.~(\ref{landaueq}) near the transition.

\begin{figure}[htbp]
\begin{center}
\includegraphics[width=7.0cm]{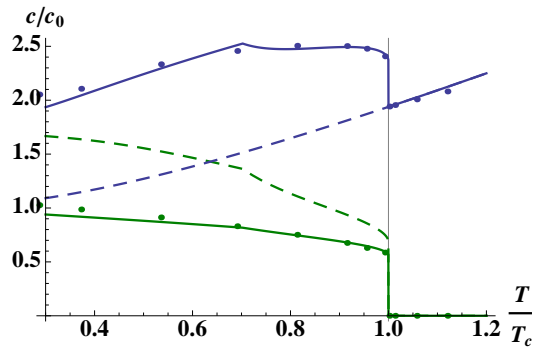}
\caption{First and second sound velocities in units of the zero temperature Bogoliubov sound velocity $c_0 = \sqrt{gn}/m$ with $g = 0.1$. The blue and green solid lines are the first and second sound velocities calculated from solving Eq.~(\ref{landaueq}). The blue and green dashed lines are the $c_{10}$ and $c_{20}$ defined in Eq.~(\ref{c10c20}). The dots are the first and second sound velocities calculated from the numerical values of~\cite{Prokof'ev2002}.}
\label{c1c2}
\end{center}
\end{figure}

In Fig.~\ref{c1c2} we show the  values for the first and second sound velocities predicted by the solutions of Eq.~(\ref{landaueq}). These values are expressed in units of the zero temperature value 
of the Bogoliubov sound velocity $c_0 \equiv \sqrt{gn}/m$ and are calculated at  fixed total density.  The most remarkable feature emerging from the figure is the discontinuity exhibited by both the first and second sound velocities at the transition.

Changing the interaction $g$ does not affect the overall qualitative behavior of the sound velocities.
For example, while for $g = 0.1$, the jumps at the transition are $c_1/c_0 = 2.36 \to 1.94$ and $c_2/c_0 = 0.56 \to 0$, reducing the value of $g$ by a factor of $3$ the jumps become  $c_1/c_0 = 3.52 \to 2.82$ and $c_2/c_0 = 0.56 \to 0$, while increasing it by a factor of $3$ the jumps become $c_1/c_0 = 1.57 \to 1.38$ and $c_2/c_0 = 0.53 \to 0$. These results show that the jump of second sound, when expressed in terms of $c_0$, is not very sensitive to the value of $g$.
Using the parameters from the experiment of~\cite{Desbuquois2012} carried out on a gas of  $^{87}$Rb atoms ($g$ = 0.093, $n$ = 50/$\mu \mathrm{m}^2$), we predict the value $c_2 \approx 0.88~\mathrm{mm/sec}$ for the second sound velocity at the transition.
This value is close to the critical velocity observed in~\cite{Desbuquois2012}, thereby suggesting that the excitation of second sound is a possible mechanism for the onset of dissipation in this experiment. 

Concerning the physical characterization of the two sounds it is worth noticing  that, according to Landau's hydrodynamic equations, the first and second sound velocities are well represented by the expressions 
\begin{align}
	c_{10} &\equiv \frac{1}{\sqrt{mn\kappa_s}},
	&
	c_{20} &\equiv \sqrt{\frac{T}{m} \frac{\bar{s}^2 n_s}{\bar{c}_p n_n}}, \label{c10c20}
\end{align}
if the conditions
\begin{align}
	\frac{c_{20}^2}{c_{10}^2}
	&\ll 1,
	&
	\frac{c_{20}^2}{c_{10}^2}
	\alpha T
	&\ll 1,
	\label{condition}
\end{align}
are both satisfied.
These formula describe accurately the solutions of the Landau's equations in the case of superfluid Helium as well as in the case of the 3D Fermi gas at unitarity.
We also expect that formula Eq.~(\ref{c10c20}) describe the sound velocities of strongly interacting 2D Fermi gases~\cite{Frohlich2011, Orel2011}.
On the other hand, in dilute Bose gases, the second condition Eq.~(\ref{condition}) is violated in a wide interval of temperatures, as a consequence of the sizable value of the expansion coefficient. For this reason, the expressions Eq.~(\ref{c10c20}) for the sound velocities turn out to be rather inaccurate [they are plotted as dashed lines in Fig.~(\ref{c1c2})].  
In particular, in dilute Bose gases, first and second sound cannot be interpreted, respectively, as isoentropic and isobaric oscillations as predicted by Eq.~(\ref{c10c20}).
This is the case of  the  weakly interacting 3D Bose gas, where the second sound speed is well approximated by the expression ~\cite{PitaevskiiStringari}
\begin{align}
	c_2 =\sqrt{ \frac{n_s}{n}\frac{1}{mn\kappa_T}}, \label{c2approx}
\end{align}
rather than by Eq.~(\ref{c10c20}), in the relevant region of temperatures $T\gg \mu$, where the isoentropic compressibility, the entropy, and the specific heat at constant volume 
are well approximated by the ideal Bose gas model, and  $n_s$ can be safely replaced by the the condensate density.
Equation (\ref{c2approx})  describes exactly the second sound velocity also in 2D Bose gases in the limit of small interactions ($g \to 0$) and, for $g = 0.1$, is a good approximation (within
$\sim 10\%$)
to the second sound  in the whole range of temperatures shown in Fig.~\ref{c1c2}.
On the other hand, the first sound velocity around the transition can be estimated by solving Eq.~(\ref{landaueq}) for small values of $g$, which gives
\begin{align}
	c_1^2 = c_{10}^2 + \alpha T c_{20}^2.
\end{align}
This result shows that both the nonzero thermal expansion coefficient and the discontinuity in the superfluid density are responsible for the jump in the first sound velocity.
Thus, the jump in the  second sound velocity is a direct consequence of the jump of the superfluid density, while the discontinuity of the first sound velocity is also the consequence of the sizable difference between the isothermal and isoentropic compressibilities discussed above.

We find that in the whole interval of temperatures, second sound in the 2D Bose gas  corresponds to an oscillation where mainly the superfluid is moving, the normal part remaining  practically at rest.  First sound, on the other hand, corresponds to an oscillation involving mainly the normal component.
We note that these features are also true in weakly interacting 3D Bose gases where actually second sound was identified as an oscillation of the condensate density \cite{Meppelink2009}. 

\begin{figure}[htbp]
\begin{center}
\includegraphics[width=7.0cm]{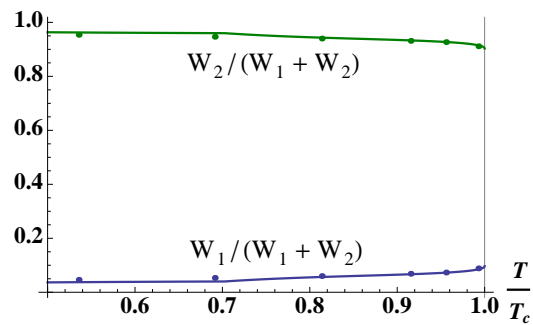}
\caption{Relative contributions of  first and second sound to the compressibility sum rule for $g = 0.1$. The bottom line corresponds to $W_1/(W_1+W_2)$ and the top line is $W_2/(W_1+W_2)$, which are the contributions from the first and the second sound, respectively. As before, the dots are calculated using the numerical values from~\cite{Prokof'ev2002}.}
\label{sumrule}
\end{center}
\end{figure}

In order to  measure  first and second sound, a first important requirement is the reachability of the collisional hydrodynamic regime of fast collisions ($\omega \tau \ll 1$, where $\omega$ is the frequency of the sound and $\tau$ is a typical collisional time) in  the normal part. 
This requirement is likely more problematic for first sound due to its higher frequency.
The excitation of second sound should be more easily accessible not only because the velocity is lower but also because in dilute  2D Bose gases it can be naturally excited by density perturbations.
For example, using a sudden laser perturbation, applied to the center of the trap, one excites both first and second sound with a relative weight given by the relative contribution of the two modes to the inverse energy weighted moment $\int_{-\infty}^\infty S(\mathbf{q},\omega)/\omega$, where $S(\mathbf{q},\omega)$ is the dynamical structure factor with momentum $\mathbf{q}$ and frequency $\omega$~\cite{Hu2010}.
This moment coincides with $n\kappa_T/2$ in the limit $q \to 0$, which is known as the compressibility sum rule~(\cite{PitaevskiiStringari}~\S 7).
Taking into account the fact that at small wave vectors not only the inverse energy weighted sum rule, but also the energy weighted  moment $\int_{-\infty}^\infty \omega S(\mathbf{q},\omega) = q^2/2m$ is exhausted by the two sound modes~\cite{NozieresPines}, one straightforwardly finds that the relative contributions to the compressibility sum rule from each sound are given by~\cite{Hu2010}
\begin{align}
	W_1 &\equiv \frac{1 - mn\kappa_T c_2^2}{c_1^2-c_2^2},
	&
	W_2 &\equiv \frac{mn\kappa_T c_1^2 - 1}{c_1^2-c_2^2}.
\end{align}
The results for the 2D Bose gas  are reported in Fig.\ref{sumrule} and show that second sound provides most of the contribution to the compressibility sum rule in the relevant temperature region $\mu \ll T < T_c$, thereby making its experimental excitation favorable through density perturbations~\footnote{
The huge contribution of second sound to the compressibility sum rule in dilute Bose gases also  explains  why, in the experiment of \cite{Meppelink2009} carried out on a 3D Bose gas, second sound could be easily excited by a density perturbation.}. At lower temperatures, in the phonon regime, the situation is modified and a typical hybridization effect between the two sounds takes place \cite{PitaevskiiStringari}.
As $T \to 0$, the thermodynamics is governed by phonons and the second sound velocity approaches the value $c_0/\sqrt{2}$.
Above $T_c$, the huge difference between the isoentropic and isothermal compressibilities is instead responsible for the occurrence of a diffusive mode at low frequency~\cite{Hu2010}.
Second sound in the 2D BKT gas could be also excited using two photon Bragg spectroscopy, a technique sensitive, at finite temperature, to the imaginary part of the density response function. 

Typically, in experiments of dilute ultracold gases, atoms are harmonically trapped.
In such systems, $T/T_c$ depends on the local density, and thus by exciting the sound modes through perturbing the center of the trap and tracing the propagation of the modes, one can reveal $T/T_c$ dependence of the sound velocities, as observed in the case of three dimensional unitary Fermi gas~\cite{Sidorenkov2013}.

\begin{acknowledgments}
We are grateful to stimulating discussions with Lev P. Pitaevskii.
Useful discussions with Yan-Hua Hou, Iacopo Carusotto, and Jean Dalibard are also acknowledged.
This work was supported by the ERC through the QGBE grant and by Provincia Autonoma di Trento.
\end{acknowledgments}

\end{document}